# Speed of sound data and acoustic virial coefficients of two binary ($N_2$ + $H_2$) mixtures at temperatures between (260 and 350) K and at pressures between (0.5 and 20) MPa

José J. Segovia[a], Daniel Lozano-Martin[a], Dirk Tuma[b], Alejandro Moreau[a], M. Carmen Martín[a], David Vega-Maza[a,*]

* Corresponding Author: david.vega@uva.es, Tel: +34 983184690

[a] TERMOCAL Research Group, Research Institute on Bioeconomy (BioEcoUVa), University of Valladolid, Escuela de Ingenierías Industriales, Paseo del Cauce 59, 47011 Valladolid, Spain.

[b] BAM Bundesanstalt für Materialforschung und -prüfung, D-12200 Berlin, Germany.

## Abstract

This work aims to address the technical concerns related to the thermodynamic characterization of gas mixtures blended with hydrogen for the implementation of hydrogen as a new energy vector. For this purpose, new experimental speed of sound measurements have been done in gaseous and supercritical phases of two binary mixtures of nitrogen and hydrogen using the most accurate technique available, i.e., the spherical acoustic resonator, yielding an experimental expanded ($k$ = 2) uncertainty of only 220 parts in $10^6$ (0.022 %). The measurements cover the pressure range between (0.5 and 20) MPa, the temperature range between (260 and 350) K, and the composition range with a nominal mole percentage of hydrogen of (5 and 10) mol %, respectively. From the speed of sound data sets, thermophysical properties that are relevant for the characterization of the mixture, namely the heat capacity ratio as perfect gas $\gamma^{pg}$, and the second $\beta_a$ and third $\gamma_a$ acoustic virial coefficients, are derived. These results are thoroughly compared and discussed with the established reference mixture models valid for mixtures of nitrogen and hydrogen, such as the AGA8-DC92 EoS, the GERG-2008 EoS, and the recently developed adaptation of the GERG-2008 EoS, here denoted GERG-$H_2$_improved EoS. Special attention has been given to the effect of hydrogen concentration on those properties, showing that only the GERG-$H_2$_improved EoS is consistent with the data sets within the experimental uncertainty in most measuring conditions.

# 1. Introduction.

The introduction of hydrogen into the existing natural gas grid is a practical alternative for the transport and storage of energy [1] produced from the surplus electricity of renewable sources, such as wind, solar, and hydraulic power [2,3], and from the gasification of organic substances by steam reforming combined with Carbon Capture, Utilization and Storage (CCUS) methods [4]. Long-term hydrogen contents around 5 % are conceivable [5], while local and temporary concentrations can reach more than 80 % close to the point of injection [6]. In particular, exploiting synergies between CCUS and the hydrogen economy could contribute, amongst other solutions, to decarbonize the energy system in Europe [7]. Design and operation of such energy systems call for an accurate characterization of the thermodynamic properties of the relevant $CO_2$- and $H_2$-mixtures.

An equation of state (EoS) specifically developed for the estimation of the thermodynamic properties of mixtures in CCUS processes is the multiparametric Helmholtz energy-based EoS-CG 2016 [8]. This EoS, however, is limited to carbon dioxide, water, nitrogen, oxygen, argon, and carbon monoxide, but an extended and by now still unpublished version, EoS-CG 2019 [9], introduces hydrogen, apart from methane, hydrogen sulfide, sulfur dioxide, hydrochloric acid, chlorine, and the two amines *N*-methylethanolamine (MEA) and diethanolamine (DEA), respectively.

Since a calculation of the thermophysical properties of a mixture composed of these substances by such an equation-of-state model depends on the composition-weighed sum of the pure-substance equations of state plus the pairwise sum of the reducing and departure functions for all the binary combinations of the considered components, the quality of the underlying binary experimental data sets limits the achievable accuracy of the equation. Within this framework, the present work deals with the assessment of three reference equations of state providing new accurate experimental data of a key thermodynamic property, the speed of sound $w$, for the binary system of nitrogen and hydrogen. This study is a continuation of a previous work at our institution where density of the same binary system was characterized [10].

With respect to the binary ($N_2 + H_2$) mixtures, the mixing rules used for the EoS-CG 2019 [9] are the same used for the GERG-2008 EoS [11,12], a model specifically developed for the description of natural gas-like mixtures. They only adjusted the parameters of the reducing functions without a departure function, since the required data sets of measured properties of binary mixtures containing hydrogen with other main components did not exist yet [12]. Regarding the speed of sound, the only data available in the literature is a limited set of low-temperature measurements reported by Van Itterbeek and Van Doninck [13], which, notably, was not implemented for the development of the GERG-2008 EoS [11,12]. Only vapor-liquid equilibria (VLE) and density data sets were used to fit the mixing parameters of the binary ($N_2 + H_2$) system in the GERG-2008 EoS [11,12], within ranges covered by experimental data between (270 and 573) K and (0.1 up to 307) MPa.

Excluding our previous work concerning the density of these mixtures [10], the experimental data available for the development of equations of state is currently restricted to hydrogen mole fractions $x_{H_2} = (0.15 \text{ to } 0.87) \text{ mol·mol}^{-1}$. This is the reason why our study focuses on mixtures of relatively small fractions of hydrogen, $x_{H_2} = (0.05 \text{ and } 0.10) \text{ mol·mol}^{-1}$. Filling this gap would be relevant for the validation and improvement of the established reference equations of state.

At the University of Bochum, Germany, recently, a new research campaign focused on four binary mixtures with hydrogen, namely $H_2 + CH_4$, $H_2 + N_2$, $H_2 + CO_2$, and $H_2 + CO$, has been carried out [14]. The newly developed EoS uses the same functional forms of the mixture models of GERG-2008 EoS [11,12], but new reducing and also specific departure functions for those four systems aforementioned have been developed. In respect of the binary ($N_2 + H_2$) mixture, this new model, named in this work as GERG-$H_2$_*improved* EoS [14], works with a broader and more accurate VLE data set, along with similar density data as those used for the GERG-2008 EoS [11,12]. This update enables the fitting of a new departure function not present in the original

GERG-2008 EoS, yielding a six-fold reduction in the overall absolute relative deviations of the new model for the phase envelope, and a twofold reduction within the homogeneous region when compared to the GERG-2008 model [11,12]. Notably, the new GERG-$H_2$_improved model [14] is already implemented in the latest version of REFPROP 10 [15] and TREND 5.0 [16] software.

The present work reports new experimental speed of sound data in gaseous and supercritical phases of the binary ($N_2$ + $H_2$) mixture. Two mixtures, with nominal mole fractions $x_{H_2}$ = (0.05 and 0.10) have been investigated, aimed at filling the gap of available data for low $H_2$ concentrations. Temperatures span between (260 and 350) K, with pressures decreasing from the highest possible value for the apparatus used in this study of 20.0 MPa down to 0.5 MPa, as illustrated in Figure 1. The measurements have then be compared with the thermodynamic mixture models widely used by the natural gas industry, such as the AGA8-DC92 EoS [17,18], the original reference GERG-2008 EoS [11,12], and the newly developed GERG-$H_2$_improved EoS [14].

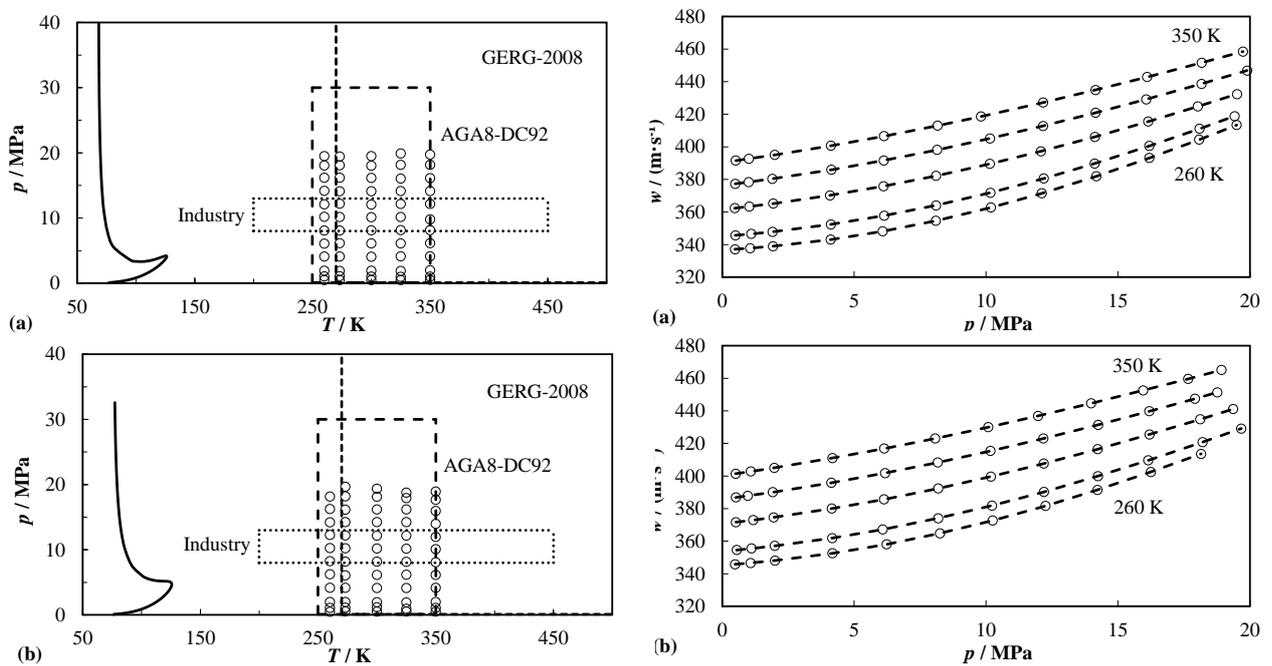

**Figure 1.** $p$, $T$ and $w$, $p$ phase diagrams showing the experimental points measured ($\bigcirc$) and the calculated phase envelope (solid line) using the GERG-2008 EoS [11,12] for: a) the (0.95 $N_2$ + 0.05 $H_2$) mixture and b) the (0.90 $N_2$ + 0.10 $H_2$) mixture. The marked temperature and pressure ranges represent the application areas covered by the AGA8-DC92 EoS [18,19] (long dashed line); the range of the binary ($N_2$ + $H_2$) experimental homogeneous data used for the development of the GERG-2008 EoS [11,12] (short dashed line); and the preferred area of interest defined by the gas industry (dotted line).

## 2. Materials and methods.

### 2.1 Mixtures.

The two binary ($N_2$ + $H_2$) test sample mixtures were provided by the German Federal Institute for Materials Research and Testing (BAM Bundesanstalt für Materialforschung und -prüfung), with the certified molar composition $x_i$ and the corresponding expanded ($k$ = 2) uncertainty $U(x_i)$ reported in Table 1. They were prepared by the gravimetric method according to ISO 6142-1 [20].

The impurities of the pure gases, such as oxygen, carbon dioxide and carbon monoxide, were considered in the uncertainty $U(x_i)$. Information on impurities were taken from the purity statements issued by the supplier, since no further purification was conducted.

**Table 1.** Purity, supplier, critical parameters of the pure components used for the preparation of the binary ($N_2 + H_2$) mixtures and mole compositions $x_i$ and expanded ($k = 2$) uncertainty $U(x_i)$ of the binary ($N_2 + H_2$) mixtures prepared at BAM and studied in this work.

| | CAS-number | Supplier | Purity in mole fraction | $M$ / g·mol$^{-1}$ | Critical parameters[a] | |
|---|---|---|---|---|---|---|
| | | | | | $T_c$ / K | $p_c$ / MPa |
| Nitrogen | 7727-37-9 | Linde AG | $\geq 0.999999$ | 28.0135 | 126.192 | 3.3958 |
| Hydrogen (normal) | 1333-74-0 | Linde AG | $\geq 0.999999$ | 2.01588 | 33.145 | 1.2964 |

| Components | (0.95 $N_2$ + 0.05 $H_2$)[b] | | (0.90 $N_2$ + 0.10 $H_2$)[c] | |
|---|---|---|---|---|
| | $10^2 \cdot x_i$ / mol·mol$^{-1}$ | $10^2 \cdot U(x_i)$ / mol·mol$^{-1}$ | $10^2 \cdot x_i$ / mol·mol$^{-1}$ | $10^2 \cdot U(x_i)$ / mol·mol$^{-1}$ |
| Nitrogen | 94.9999 | 0.0013 | 90.0005 | 0.0013 |
| Hydrogen (normal) | 5.0001 | 0.0016 | 9.9995 | 0.0027 |

[a] Critical parameters were obtained by using the default equation of state for each substance in REFPROP software [15], i.e., the reference equation of state for nitrogen [21] and the reference equation of state for hydrogen (normal) [22].

[b] BAM cylinder no.: 96054 968-160517 (G 020).

[c] BAM cylinder no.: 96054 970-160501 (G 020).

These two mixtures, i.e., the same cylinders, were previously used to measure the density in the gaseous and supercritical states with the results reported in ref. [10], where more technical details about the filling procedures, the apparatus used for the mixture preparation, and gas chromatography (GC) validation results can be found. The GC analysis of the mixtures carried out at BAM to validate the mixture according to ISO 12963 [23], reveals that the maximum relative deviation of $x_{N_2}$ is less than 0.006 % and that of $x_{H_2}$ is less than 0.072 %, within the expanded ($k = 2$) relative uncertainty of the GC validation, i.e., $U_{r,GC}(x_{N_2}) = 0.020$ % and $U_{r,GC}(x_{H_2}) = 0.048$ %, respectively. Prior to measuring the speed of sound, the cylinders containing the prepared mixtures were homogenized by rolling and heating for several hours.

### 2.2 Experimental setup.

The speed of sound was measured using a spherical stainless-steel acoustic resonator designed for measurements of pure gases and their mixtures. A detailed description of the components which comprise the experimental setup has been given in previous works [24–27]. The resonator was used in this research without further modifications. In brief, the main part of the acoustic resonator consists of a spherical stainless-steel A321 acoustic cavity with a nominal radius of $40 \cdot 10^{-3}$ m and a wall thickness of $12.5 \cdot 10^{-3}$ m, respectively. The design of the resonator originates from the works of Ewing and Trusler [28,29] and it was assembled at Imperial College London workshops meeting the necessary tolerances and roughness requirements for its application. The resonant cavity serves as a pressure-tight shell, with two ports to accommodate the capacitance microphones and two additional ducts of radius $r_0 = 0.8 \cdot 10^{-3}$ m and lengths $L_1 = 2.3$ m and $L_2 = 0.035$ m, respectively. The former duct serves as inlet gas tube, whereas the latter was of no use in the present work, remaining closed by means of a plug. The determination of the variation of its internal radius with the temperature and pressure has been accomplished and verified elsewhere [26,30], by acoustic calibration in argon, inasmuch as its equation of state is well-known [31], with a low expanded ($k = 2$) relative uncertainty in the calculated speed of sound that amounts to 0.02 %.

Two specially designed and identical acoustic capacitance transducers are embedded flush with the inner wall of the resonant cavity, forming an angle of 90º between them to reduce overlapping of the non-degenerate modes with the close degenerate ones. They are made of polyimide film of 12 μm thickness with an active area of 3 mm diameter and are gold-plated in the side facing the resonance cavity. The source transducer is driven without any bias voltage by a wave generator (3225B function generator, HP) at a peak-to-peak signal of 40 V amplified up to 180 V, causing it to vibrate at twice the selected frequency in the wave synthesizer and thus reducing the crosstalk with the detector transducer. A lock-in amplifier (SR850 DSP Lock-In, SRS) measures the driven signal by triaxial cables from the detector transducer, fed from an external amplifier which maintains the detector polarized with a DC signal of 80 V. The precision in the frequency measurement $f$ is better than $10^{-7} \cdot f$. More details about the instrumentation and procedures used for the acquisition and fitting of the acoustic signal to determine the experimental acoustic resonance frequencies $f_{0n}$ and halfwidths $g_{0n}$ of the purely radial ($l = 0,n$) acoustic standing-wave modes are given in previous works [32,33].

The acoustic cavity is contained within a stainless-steel vacuum vessel, which in turn is submerged in an ethanol-filled Dewar vessel cooled by a thermostat (FP89-ME, Julabo). The temperature control of the vessel at the selected set point is achieved by three temperature loops encompassing band resistors and 25-Ω standard platinum resistance thermometers (SPRTs). The latter are located at the base and side of the vessel, and at the top copper block from which the acoustic cavity is suspended. The observed thermal gradient between hemispheres does not exceed a value of 1 mK, with an achievable better temperature stability. Two standard platinum resistance thermometers (25.5 Ω SPRT 162D, Rosemount) that have been calibrated against an ITS-90 and located in the northern and southern hemispheres are plugged into an AC resistance bridge (F18 automatic bridge, ASL) to measure the temperature of the gas inside with an expanded ($k = 2$) uncertainty from calibration of $U(T) = 4$ mK.

The pressure inside of the resonant shell is monitored by two piezoelectric quartz transducers located at the top of the inlet gas tube: one calibrated for the pressure range (0 to 2) MPa, (2003A-101-CE, Paroscientific Digiquartz); and the other for the pressure range (2 to 20) MPa, (43KR-101-CE, Paroscientific Digiquartz). The pressure values include the hydrostatic pressure correction. The expanded ($k = 2$) uncertainty from their calibration is estimated to be $U(p) = (8 \cdot 10^{-5} \, (p/\text{Pa}) + 200)$ Pa.

### 2.3 Data analysis.

The speed of sound for each radial acoustic mode $w_{0n}$ is obtained from the mean of 5 repetitions of $f_{0n}$ and $g_{0n}$ at each thermodynamic state ($p_i$, $T_{i,\text{exp}}$), after correcting to the same reference temperature $T_{\text{ref}}$ by multiplying them by the ratio of the estimated speed of sound values from REFPROP 10 [15] $w(p_i, T_{\text{ref}})$ / $w(p_i, T_{i,\text{exp}})$, following:

$$w_{0n}(p_i, T_{\text{ref}}) = 2\pi a(p_i, T_{\text{ref}}) \frac{(f_{0n} - \Delta f_{\text{th}} - \Delta f_{\text{sh}} - \Delta f_{\text{tr}} - \Delta f_0)}{z_{0n}} \tag{1}$$

where $a(p_i, T_{\text{ref}})$ stands for the internal radius of the resonance cavity, $z_{0n}$ stands for the $n$-th zero of the spherical Bessel function first derivative of order $l = 0$, and $\Delta f_{\text{th}}$, $\Delta f_{\text{sh}}$, $\Delta f_{\text{tr}}$, and $\Delta f_0$ stand for the corrections of the thermal boundary layer [34], coupling of fluid and shell motion [35], transducer perturbation [36], and gas ducts perturbation [37], respectively.

The full set of expressions used to evaluate these perturbations to $f_{0n}$ is provided elsewhere [38,39]. The thermodynamic and transport properties of the gas mixture required by these corrections have been computed from REFPROP 10 [15] using the GERG-2008 EoS [11,12], whereas the elastic and thermal properties of the cavity material have been obtained from the models and published values of the properties of stainless steel A321 [40–42].

The overall magnitude of the relative frequency corrections $\Delta f / f_{0n} = (\Delta f_{th} + \Delta f_{sh} + \Delta f_{tr} + \Delta f_0)/f_{0n}$ extends from 190 parts in $10^6$ for the (0,2) mode of the binary (0.95 $N_2$ + 0.05 $H_2$) mixture at $T = 260$ K and the lowest pressure, and 560 parts in $10^6$ for the (0,4) mode of the binary (0.90 $N_2$ + 0.10 $H_2$) mixture at $T = 350$ K and the highest pressure. Hence these values are in the same order of magnitude or higher than the expanded ($k = 2$) uncertainty of the speed of sound $U_r(w_{exp})$ equal to 220 parts in $10^6$, they cannot be neglected.

The coherence of the above applied acoustic model is evaluated by means of two quantities:
1) the relative speed of sound dispersion between modes around the mean $<w>$, $\Delta w/w = (w_{0n} - <w>)/<w>$;
2) the excess relative resonance halfwidths, defined as the difference between the experimental and the acoustic model halfwidths divided by the experimental acoustic resonance frequencies: $\Delta g/f_{0n} = (g_{0n} - g_{th} - g_0 - g_{bulk})/f_{0n}$, where $g_{th}$ stands for the thermal energy losses in the thermal boundary layer, $g_0$ stands for the losses in the gas ducts, and $g_{bulk}$ stands for the viscous and thermal losses in the bulk of the fluid.

Figures 2 and 3 depict $\Delta g/f_{0n}$ and $\Delta w/w$, respectively, as a function of pressure for the two mixtures studied in this work at the intermediate temperature $T = 300$ K. In Figure 3, a disagreement between the (0,6) mode for the binary (0.95 $N_2$ + 0.05 $H_2$) and (0.90 $N_2$ + 0.10 $H_2$) mixtures, as well as the (0,5) mode for the binary (0.90 $N_2$ + 0.10 $H_2$) mixture, compared to the others is shown. In addition, Figure 2 shows that the $\Delta g/f_{0n}$ values for the mentioned cases are larger than $U_r(w_{exp}) = 220$ parts in $10^6$ at the highest pressures for the (0,5) mode and over the entire pressures investigated for the (0,6) mode, indicating that the current acoustic model fails to fully reproduce the physical situation within the acoustic cavity. Thus, these modes are discarded from any following calculation at this temperature.

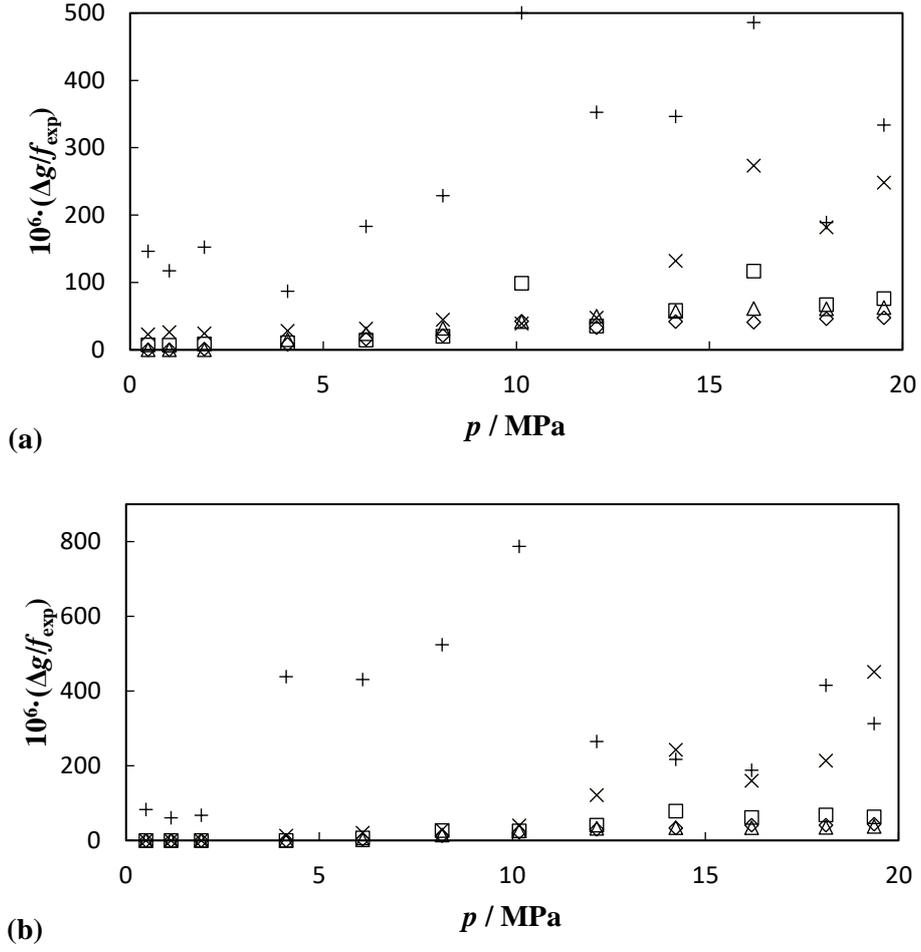

**Figure 2.** Relative excess halfwidths ($\Delta g/f_{\text{exp}}$) for (a) the mixture (0.95 N$_2$ + 0.05 H$_2$) and (b) the mixture (0.90 N$_2$ + 0.10 H$_2$), at $T$ = 300 K and modes $\triangle$ (0,2), $\diamondsuit$ (0,3), $\square$ (0,4), $\times$ (0,5), $+$ (0,6), previously to apply the vibrational relaxation correction.

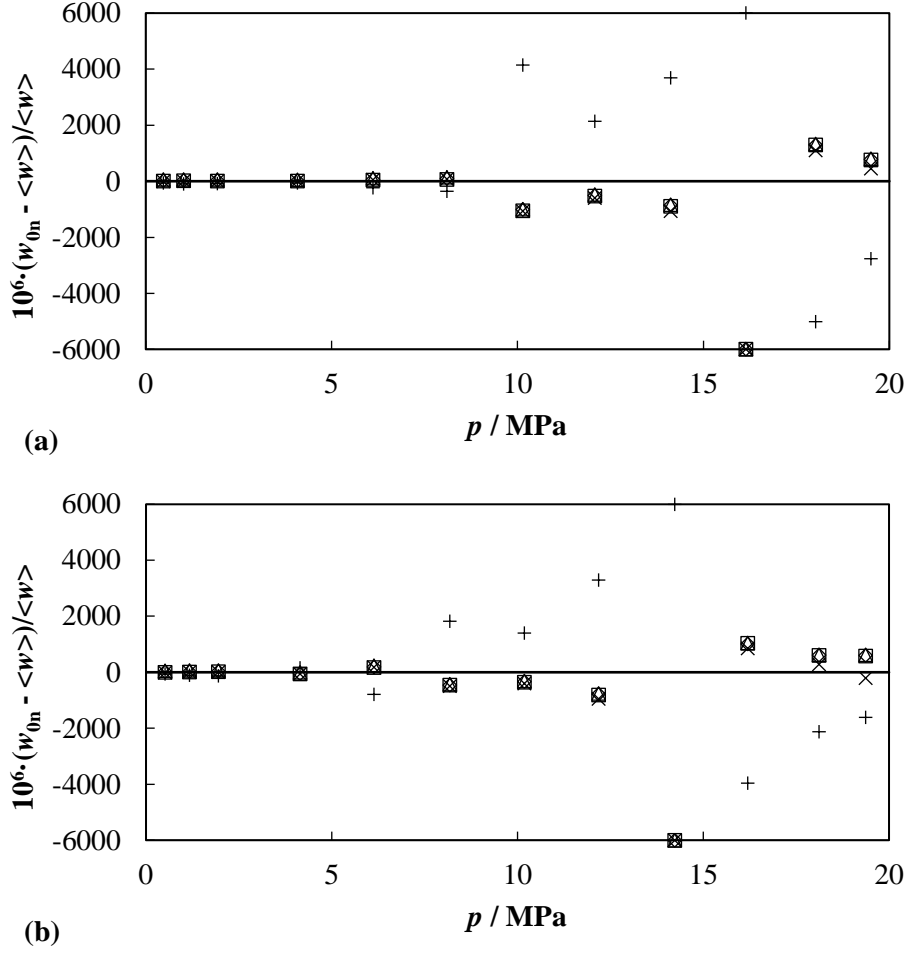

**Figure 3.** Relative dispersion of the speed of sound $\Delta w/w = (w_{0n} - <w>)/<w>$, where $<w>$ is the mean value for (0,2) to (0,6) radial acoustic modes, as a function of pressure $p$ at temperature $T = 300$ K for (a) the mixture (0.95 N$_2$ + 0.05 H$_2$) and (b) the mixture (0.90 N$_2$ + 0.10 H$_2$). Modes: △ (0,2), ◇ (0,3), □ (0,4), × (0,5), + (0,6).

A similar analysis was conducted for the other isotherms, leading to neglect the (0,5) and (0,6) modes in all cases, with the exception at $T = (260$ to $300)$ K for the (0.95 N$_2$ + 0.05 H$_2$) mixture and at $T = 260$ K for the (0.90 N$_2$ + 0.10 H$_2$) mixture, respectively, where the (0,5) modes were additionally included. Note that the (0,6) mode has not been measured at temperatures higher than 300 K, because the resonance frequencies fall too close to the mechanical resonance frequency of the transducers, which amounts to about 40 kHz, thus discouraging us from their acquisition to avoid damage and masking the fluid response. For the remaining (0,2), (0,3), and (0,4) modes, the mean relative dispersion of the speed of sound data is in the order of 40 parts in $10^6$ between them, and the corresponding excess halfwidths are always less than 60 parts in $10^6$ for these modes, showing a fairly constant behavior with the temperature and the pressure, unlike the excluded (0,5) and (0,6) modes mentioned before.

The vibrational relaxation time of pure nitrogen is known to be very long [43,44] and, on first approximation, should be considered as infinite. Looking at the literature related to the speed of sound in pure hydrogen [45–50], there is no evidence of sound dispersion reported, neither in the range of temperatures of this work [46,51] . Thus, on first approximation, hydrogen should be taken as a non-relaxing gas. In this situation there is only one fast vibrational relaxation time associated with the translational and rotational relaxation of nitrogen molecules colliding with hydrogen molecules $\tau_{vib} = \tau_{N_2\text{-}H_2}/x_{H_2}$. Assuming that $\tau_{N_2\text{-}H_2}\rho_n$ is constant along an isotherm, the worst-case scenario is at $T = 350$ K for the mixture (0.95 N$_2$ + 0.05 H$_2$) and the non-discarded mode of highest frequency, the (0,4). The vibrational relaxation times derived from the corresponding excess halfwidths $\tau_{vib}$ are [36]:

$$\tau_{vib} = \frac{\Delta g_{0n}}{\Delta(\gamma-1)\pi f_{0n}^2} \tag{2}$$

with $\Delta$ the vibrational contribution to the isobaric heat capacity of the mixture:

$$\Delta = \sum_k \frac{C_{vib,k}}{C_P M} \tag{3}$$

where the vibrational heat capacities are evaluated from Plank-Einstein functions and vibrational frequencies from spectroscopy $\nu_i$ [52]:

$$C_{vib,k} = R\sum_k \frac{z_i^2 e^{z_i}}{\left(e^{z_i}-1\right)} \tag{4}$$

$$z_i = \frac{\vartheta_i}{T} = \frac{h_P \nu_i / k_B}{T} \tag{5}$$

where $h_P$ stands for the Planck constant and $k_B$ for the Boltzmann constant. The vibrational correction is then given by:

$$\frac{\Delta f_{vib}}{f_{0n}} = \frac{1}{2}(\gamma-1)\Delta\left(2\pi f_{0n}\tau_{vib}\right)^2\left(1-\frac{\Delta(1+3\gamma)}{4}\right) \tag{6}$$

The resulting $\tau_{vib}$ are within (10 to 75)$\cdot 10^{-6}$ s with corresponding relative vibrational corrections $\Delta f_{vib}/f_{0n}$ ranging from (1 to 83) part in $10^6$ for the pressure range from (0.5 to 20) MPa, with lower values in the other scenarios. We conclude the vibrational correction must be applied to all data points. At the point of greatest value of $\Delta g/f_{0n}$, measured for the mixture of higher nitrogen content at the highest temperature and highest not neglected resonance mode, the mode (0,4) with resonance frequencies between (1.7 to 1.95)$\cdot 10^4$ Hz: (i) the relative excess halfwidths of mode (0,4) are reduced from a maximum value of (169 to 49) parts in $10^6$ after allowing for vibrational relaxation, and (ii) the remaining excess halfwidth contributes to a reduced experimental expanded ($k$ = 2) uncertainty of 240 parts in $10^6$. This approximate model assumes that all the molecules relax at unison, including the relaxation for both nitrogen and hydrogen.

A possible instability of the mixture due to a higher adsorption in the cavity walls of one of the components compared to the other, resulting in a corresponding change in the composition of the mixture, was assessed by continuous acoustic frequency measurements at the same state point. As shown in Figure 4, the tests were performed for the acoustic (0,3) mode at the bottle pressure after finishing all the measures, $p \sim 6.6$ MPa, and the intermediate temperature $T$ = 300 K for the mixture (0.90 $N_2$ + 0.10 $H_2$). The change of the acoustic frequency was lower than 0.63 Hz (i.e., 53 parts in $10^6$) over a time of six days and equivalent to the residence time of the mixture in the resonator during the measurement of an entire isotherm. This effect, propagated in the expanded ($k$ = 2) relative uncertainty of 220 parts in $10^6$ in the experimental speed of sound, means an overall $U_r(w_{exp})$ of 225 parts in $10^6$ at maximum. A minor adsorption phenomenon is expected at higher temperatures and lower hydrogen content. Additionally, to minimize the sorption effects during the measurement, the resonance cavity was evacuated and flushed several times with fresh sample from the gas cylinders after finishing the measurements at each isotherm.

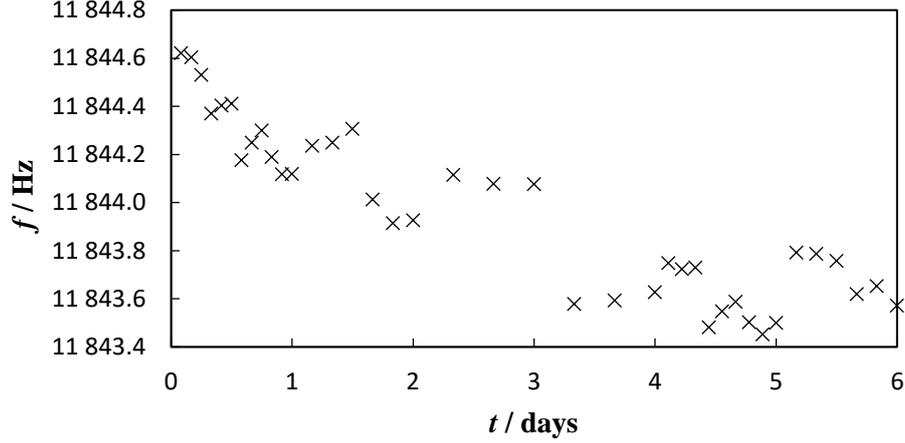

**Figure 4.** Acoustic resonance frequency as function of time (in days) corresponding to the radial acoustic (0,3) mode at $p \sim 6.6$ MPa and $T = 300$ K for the binary (0.90 $N_2$ + 0.10 $H_2$) mixture.

## 3. Results and uncertainty.

For the five isotherms $T = (260, 273.16, 300, 325, \text{ and } 350)$ K and at pressures between $p = (0.5 \text{ and } 20)$ MPa, the speed of sound data have been calculated for two binary ($N_2 + H_2$) mixtures with nominal hydrogen mole fractions $x_{H_2} = (0.05 \text{ and } 0.10)$ according to the Equation (1), considering the first five (0,2), (0,3), (0,4), (0,5), and (0,6) radial acoustic modes. Following the non-consideration of some of the (0,5) and (0,6) acoustic modes as discussed above, average single estimates for the speed of sound $w_{exp}(p,T)$ were calculated in the gaseous and supercritical homogeneous regions. Tables 2 and 3 show these data sets for the two binary mixtures of the composition (0.95 $N_2$ + 0.05 $H_2$) and (0.90 $N_2$ + 0.10 $H_2$) mixtures, respectively.

**Table 2.** Speed of sound $w_{exp}(p,T)$ for the mixture (0.95 $N_2$ + 0.05 $H_2$) with their relative expanded ($k = 2$) uncertainty[(*)] and relative deviations $(w_{exp} - w_{EoS})/w_{EoS} = \Delta w_{r,EoS}$ from the speed of sound predicted by the AGA8-DC92 EoS [18,19], GERG-2008 EoS [11,12], and GERG-$H_2\_improved$ [12]. The mixture mole fractions with their uncertainties from gravimetric preparation are provided in Table 1.

| $p$ / MPa | $w_{exp}$ / $\text{m} \cdot \text{s}^{-1}$ | $10^6 \, \Delta w_{r,AGA}$ | $10^6 \cdot \Delta w_{r,\text{GERG-2008}}$ | $10^6 \, \Delta w_{r,\text{GERG-}H2\_improved}$ |
|---|---|---|---|---|
| | | $T = 260.000$ K | | |
| 0.47626 | 337.036 | −117 | −217 | −220 |
| 1.05689 | 337.763 | 86 | −39 | −54 |
| 1.92228 | 338.984 | 206 | 68 | 20 |
| 4.10306 | 343.072 | 304 | 217 | 40 |
| 6.07642 | 348.114 | 356 | 329 | 19 |
| 8.08968 | 354.594 | 429 | 411 | −13 |
| 10.18226 | 362.732 | 513 | 458 | −46 |
| 12.11015 | 371.428 | 576 | 472 | −71 |
| 14.17767 | 381.911 | 580 | 450 | −107 |
| 16.18919 | 393.129 | 546 | 426 | −131 |
| 18.07550 | 404.373 | 380 | 288 | −267 |
| 19.49785 | 413.099 | −122 | −199 | −752 |

$T = 273.160$ K

| | | | | |
|---|---|---|---|---|
| 0.49452 | 345.624 | −7 | −84 | −92 |
| 1.09722 | 346.517 | 137 | 34 | 8 |
| 1.90258 | 347.838 | 224 | 111 | 51 |
| 4.11145 | 352.367 | 315 | 259 | 67 |
| 6.13508 | 357.724 | 333 | 354 | 31 |
| 8.09608 | 364.029 | 368 | 433 | 3 |
| 10.15897 | 371.808 | 423 | 493 | −15 |
| 12.18647 | 380.535 | 481 | 533 | −14 |
| 14.11773 | 389.739 | 461 | 505 | −53 |
| 16.18551 | 400.502 | 498 | 559 | 9 |
| 18.09480 | 410.961 | 86 | 189 | −346 |
| 19.42026 | 418.620 | −32 | 103 | −421 |

$T = 300.000$ K

| | | | | |
|---|---|---|---|---|
| 0.47093 | 362.335 | 71 | 24 | 11 |
| 1.01972 | 363.356 | 186 | 110 | 76 |
| 1.92835 | 365.168 | 286 | 188 | 112 |
| 4.07912 | 370.156 | 357 | 314 | 112 |
| 6.11252 | 375.790 | 349 | 411 | 87 |
| 8.10013 | 382.142 | 323 | 488 | 61 |
| 10.14112 | 389.499 | 310 | 551 | 50 |
| 12.07728 | 397.218 | 320 | 613 | 72 |
| 14.12529 | 406.068 | 257 | 596 | 42 |
| 16.14441 | 415.418 | 131 | 525 | −19 |
| 18.02240 | 424.700 | 187 | 647 | 124 |
| 19.51014 | 432.282 | 50 | 569 | 69 |

$T = 325.000$ K

| | | | | |
|---|---|---|---|---|
| 0.49059 | 377.247 | 151 | 109 | 92 |
| 0.99671 | 378.321 | 251 | 177 | 139 |
| 1.89349 | 380.327 | 371 | 271 | 190 |
| 4.13475 | 385.930 | 474 | 420 | 215 |
| 6.11588 | 391.603 | 483 | 545 | 229 |
| 8.14258 | 398.068 | 432 | 634 | 222 |
| 10.14730 | 405.099 | 380 | 714 | 231 |
| 12.15918 | 412.739 | 310 | 761 | 235 |
| 14.14486 | 420.833 | 288 | 841 | 299 |
| 16.06626 | 429.116 | 240 | 895 | 357 |
| 18.16291 | 438.599 | 151 | 920 | 405 |
| 19.89374 | 446.743 | 69 | 939 | 453 |

$T = 350.000$ K

| | | | | |
|---|---|---|---|---|
| 0.50888 | 391.524 | 166 | 120 | 100 |
| 1.00493 | 392.683 | 288 | 206 | 165 |
| 1.98382 | 395.044 | 408 | 292 | 205 |
| 4.11419 | 400.651 | 534 | 422 | 256 |
| 6.12893 | 406.514 | 507 | 552 | 251 |
| 8.15527 | 412.957 | 446 | 657 | 266 |
| 9.80520 | 418.575 | 363 | 719 | 271 |
| 12.15727 | 427.136 | 253 | 808 | 306 |
| 14.14644 | 434.832 | 143 | 860 | 338 |

| p / MPa | $w_{exp}$ / m·s$^{-1}$ | | | |
|---|---|---|---|---|
| 16.10280 | 442.777 | 38 | 911 | 389 |
| 18.17671 | 451.556 | −86 | 948 | 444 |
| 19.74181 | 458.402 | −181 | 978 | 496 |

**Table 3.** Speed of sound $w_{exp}(p,T)$ for the mixture (0.90 N$_2$ + 0.10 H$_2$) with their relative expanded ($k = 2$) uncertainty$^{(*)}$ and relative deviations $(w_{exp} - w_{EoS})/w_{EoS} = \Delta w_{r,EoS}$ from the speed of sound predicted by the AGA8-DC92 EoS [18,19], GERG-2008 EoS [11,12], and GERG-$H_2\_improved$ [12]. The mixture mole fractions with their uncertainties from gravimetric preparation are provided in Table 1.

| p / MPa | $w_{exp}$ / m·s$^{-1}$ | $10^6 \, \Delta w_{r,AGA}$ | $10^6 \cdot \Delta w_{r,GERG\text{-}2008}$ | $10^6 \, \Delta w_{r,GERG\text{-}H2\_improved}$ |
|---|---|---|---|---|
| | | | $T = 260.000$ K | |
| 0.48063 | 345.730 | 26 | −152 | −160 |
| 1.07200 | 346.571 | 214 | 4 | −28 |
| 1.99069 | 348.045 | 377 | 150 | 54 |
| 4.17539 | 352.529 | 553 | 381 | 58 |
| 6.22560 | 358.058 | 648 | 526 | −26 |
| 8.25071 | 364.789 | 741 | 612 | −122 |
| 10.24503 | 372.617 | 835 | 643 | −203 |
| 12.25379 | 381.630 | 902 | 624 | −269 |
| 14.22334 | 391.453 | 872 | 533 | −360 |
| 16.24325 | 402.274 | 383 | 26 | −839 |
| 18.15044 | 413.550 | 728 | 389 | −442 |
| | | | $T = 273.160$ K | |
| 0.54515 | 354.532 | −23 | −153 | −171 |
| 1.10720 | 355.476 | 163 | 2 | −47 |
| 1.98348 | 357.073 | 311 | 137 | 21 |
| 4.16731 | 361.894 | 473 | 355 | 11 |
| 6.07628 | 367.138 | 537 | 493 | −60 |
| 8.17968 | 374.026 | 597 | 595 | −146 |
| 10.22465 | 381.800 | 660 | 647 | −212 |
| 12.19049 | 390.195 | 664 | 615 | −293 |
| 14.23909 | 399.882 | 748 | 669 | −238 |
| 16.14452 | 409.589 | 704 | 625 | −249 |
| 18.21018 | 420.811 | 647 | 606 | −215 |
| 19.67149 | 429.101 | 553 | 556 | −224 |
| | | $T = 300.000$ K | | |
| 0.51521 | 371.621 | 27 | −46 | −73 |
| 1.16761 | 372.942 | 188 | 80 | 7 |
| 1.94772 | 374.621 | 312 | 187 | 49 |
| 4.14734 | 379.997 | 466 | 398 | 38 |
| 6.13086 | 385.672 | 490 | 528 | −34 |
| 8.17959 | 392.348 | 492 | 636 | −99 |
| 10.18170 | 399.628 | 476 | 700 | −152 |
| 12.18047 | 407.598 | 464 | 740 | −171 |
| 14.23078 | 416.441 | 437 | 764 | −155 |
| 16.19178 | 425.465 | 390 | 772 | −116 |
| 18.11659 | 434.794 | 308 | 766 | −66 |
| 19.36352 | 441.055 | 236 | 752 | −35 |
| | | $T = 325.000$ K | | |
| 0.51195 | 386.820 | −18 | −19 | −52 |
| 0.95885 | 387.835 | 88 | 89 | 23 |
| 1.91421 | 390.072 | 200 | 201 | 56 |
| 4.12851 | 395.831 | 435 | 435 | 79 |
| 6.16763 | 401.806 | 565 | 566 | 17 |
| 8.16380 | 408.276 | 674 | 673 | −34 |
| 10.16534 | 415.334 | 730 | 729 | −92 |
| 12.17067 | 422.965 | 800 | 799 | −87 |
| 14.24793 | 431.390 | 838 | 837 | −69 |
| 16.18913 | 439.701 | 860 | 860 | −26 |
| 17.91437 | 447.392 | 846 | 845 | 2 |
| 18.76648 | 451.294 | 845 | 844 | 28 |
| | | $T = 350.000$ K | | |
| 0.50271 | 401.347 | −25 | −81 | −116 |
| 1.07896 | 402.770 | 169 | 72 | −7 |
| 1.97202 | 405.015 | 309 | 180 | 27 |
| 4.17342 | 410.988 | 472 | 388 | 38 |
| 6.13943 | 416.824 | 439 | 488 | −32 |
| 8.06613 | 423.007 | 357 | 577 | −85 |
| 10.08153 | 429.930 | 230 | 643 | −132 |
| 11.96862 | 436.815 | 112 | 703 | −139 |
| 13.97746 | 444.532 | −25 | 751 | −122 |
| 15.95305 | 452.472 | −179 | 776 | −91 |
| 17.65460 | 459.561 | −328 | 782 | −57 |
| 18.92092 | 465.002 | −387 | 841 | 36 |

[*] Expanded uncertainties ($k = 2$): $U(p) = (8 \cdot 10^{-5}\ (p/\text{Pa}) + 200)$ Pa; $U(T) = 4$ mK; $U_r(w) = 2.5 \cdot 10^{-4}$ m·s$^{-1}$/ (m·s$^{-1}$).

Table 4 reports the specific uncertainty contributions to the speed of sound uncertainty, characterizing all speed of sound datasets. These contributions encompass the imperfect determination of pressure and temperature, the amounts of substances of the mixtures, the calibration of the internal radius of the cavity, and the goodness of fit used to measure the frequencies and halfwidths of each radial acoustic resonance. The two indicators of the non-adequateness of the aforementioned acoustic model, the speed of sound dispersion between modes and the relative excess halfwidths, are also added to the speed of sound uncertainty. The quadrature sum of all these uncorrelated components [53] amounts to an overall relative expanded ($k = 2$) uncertainty of the speed of sound $U_r(w_{exp})$ equal to 220 parts in $10^6$ (0.022 %). As expected, the main contribution to $U_r(w_{exp})$ originates from the internal cavity radius determination as a function of pressure and temperature by the speed of sound measurements in argon, $U_r(\text{Ar})$, estimated to be up to 200 parts in $10^6$. Next in amount is the contribution of the relative excess halfwidths, $U_r(\Delta g/f_{0n})$, which amounts up to 100 parts in $10^6$. It is then followed by the (60 and 50) parts in $10^6$ from the standard deviations of the speed of sound considering the different non-rejected modes, $U(<w>)$, and the uncertainty of the gas molar mass according to the composition of the gas mixtures given in Table 1, $U_r(M)$, respectively. Minor terms influencing the uncertainty are: (i) the (12 and 8) parts in $10^6$ contributions from the temperature $U_r(T)$ and pressure $U_r(p)$ relative uncertainties, respectively, mainly due to the calibration of the corresponding probes; and (ii) the error associated to the acquisition of the resonance frequency from the detected acoustic signal by the lock-in amplifier, $U_r(f_{0n})$ of less than 4 parts in $10^6$ [40,41].

**Table 4.** Uncertainty budget for the speed of sound measurements $w_{exp}$.

| Source | Magnitude | | Contribution to the speed of sound uncertainty, $10^6 \cdot u_r(w_{exp})$ |
|---|---|---|---|
| Temperature | Calibration | $2 \cdot 10^{-3}$ K | |
| | Resolution | $7 \cdot 10^{-7}$ K | |
| | Repeatability | $5 \cdot 10^{-5}$ K | |
| | Gradient (across hemispheres) | $1.5 \cdot 10^{-3}$ K | |
| | Summation in quadrature | $3 \cdot 10^{-3}$ K | 6 |
| Pressure | Calibration | $(4 \cdot 10^{-5} \cdot p$ / MPa $+ 1 \cdot 10^{-4})$ MPa | |
| | Resolution | $3 \cdot 10^{-5}$ MPa | |
| | Repeatability | $1.1 \cdot 10^{-5}$ MPa | |
| | Summation in quadrature | $(1.2$ to $8) \cdot 10^{-4}$ MPa | 4 |
| Gas composition | Purity | $3 \cdot 10^{-7}$ kg·mol$^{-1}$ | |
| | Molar mass | $1.1 \cdot 10^{-6}$ kg·mol$^{-1}$ | |
| | Summation in quadrature | $1.1 \cdot 10^{-6}$ kg·mol$^{-1}$ | 23 |
| Radius from speed of sound in Ar | Temperature | $2 \cdot 10^{-9}$ m | |
| | Pressure | $2 \cdot 10^{-10}$ m | |
| | Gas composition | $4 \cdot 10^{-9}$ m | |
| | Frequency fitting | $5 \cdot 10^{-7}$ m | |
| | Regression | $1.7 \cdot 10^{-6}$ m | |
| | Equation of State | $2 \cdot 10^{-6}$ m | |
| | Dispersion of modes | $3 \cdot 10^{-6}$ m | |
| | Summation in quadrature | $4 \cdot 10^{-6}$ m | 100 |
| Frequency fitting | | $2.3 \cdot 10^{-2}$ Hz | 2 |
| Dispersion of modes | | $1.5 \cdot 10^{-2}$ m·s$^{-1}$ | 30 |
| Relative excess halfwidths after allowing for vibrational correction | | | 50 |
| Summation in quadrature of all contributions to $w_{exp}$ | | | 110 |
| Relative expanded uncertainty ($k = 2$): | | $10^6 \cdot U_r(w_{exp})$ | 220 |

Squared speed of sound data $w^2_{exp}(p_i,T_{ref},x)$ where $x$ represents the nominal mole fraction of hydrogen in the mixture, are fitted to a power series expansion in the pressure $p$, denoted as the acoustic virial equation:

$$w^2_{exp}\left(p,T_{ref},x\right) = A_0\left(T_{ref},x\right) + A_1\left(T_{ref},x\right)p + A_2\left(T_{ref},x\right)p^2 + ... \quad (7)$$

from which: (i) the heat capacity ratio as perfect gas $\gamma^{pg}$ is obtained as $\gamma^{pg} = M \cdot A_0 / (R \cdot T)$, (ii) the second acoustic virial coefficient $\beta_a$ is determined as $\beta_a = R \cdot T \cdot A_1 / A_0$; and (iii) the third acoustic virial coefficient $\gamma_a$ is derived as $\gamma_a = (R \cdot T)^2 \cdot A_2 / A_0 + \beta_a \cdot B(T,x)$. $M$ stands for the gas molar mass, $R$ for the molar gas constant, and $B(T,x)$ for the second density virial coefficient. The degree of Equation (7) is chosen according to the criteria that 1) the residuals of the fitting are randomly distributed within the expanded ($k = 2$) uncertainty of the speed of sound $U(w_{exp})$, and 2) the computed uncertainties of the regression parameters by the Monte Carlo method [54] are lower than their own magnitudes. Both criteria determine that all the regression parameters $A_i$, which are given in Table 5 together with their estimated expanded ($k = 2$) uncertainties, are significant for the chosen degree. As shown in Table 5, a fourth-order polynomial of Equation (7) is required at $T = (260$ and $273.16)$ K for the binary $(0.95 \, N_2 + 0.05 \, H_2)$ and $(0.90 \, N_2 + 0.10 \, H_2)$ mixtures, while a second-order is sufficient for the remaining isotherms. The relative root mean squares of the residuals of the fitting are lower than 150 parts in $10^6$ for all the temperatures, which is well within $U_r(w_{exp})$, as it is shown in Figure 5. Table 6 reports the results for $\gamma^{pg}$, $C^{pg}_{p,m}$, $\beta_a$, and $\gamma_a$ obtained from the fitting parameters $A_i$ reported in Table 5, together with their expanded ($k = 2$) uncertainties. The latter consider the uncertainties of the fitting parameters $U(A_i)$ computed with the Monte Carlo method [54], the temperature uncertainties $U(T_{ref})$, and the molar mass of the mixtures uncertainties $U(M)$. Note that, the uncertainty of the molar gas constant $R$ is now zero after the redefinition of the kelvin in 2019 [25,55].

**Table 5.** Fitting parameters $A_i(T)$ to the squared speed of sound-pressure data with the equation (2) for each mixture, their corresponding expanded ($k = 2$) uncertainties determined by the Monte Carlo method [54], and the root mean square ($\Delta_{RMS}$ = root mean square relative deviations) of the residuals of the fitting.

| $T$ / K | $A_0(T)$ / (m²·s⁻²) | $10^6 \cdot A_1(T)$ / (m²·s⁻²·Pa⁻¹) | $10^{12} \cdot A_2(T)$ / (m²·s⁻²·Pa⁻²) | $10^{19} \cdot A_3(T)$ / (m²·s⁻²·Pa⁻³) | $10^{27} \cdot A_4(T)$ / (m²·s⁻²·Pa⁻⁴) | $\Delta_{RMS}$ of residuals / ppm |
|---|---|---|---|---|---|---|
| | | | $(0.95 \, N_2 + 0.05 \, H_2)$ | | | |
| 260.000 | $113259 \pm 11$ | $678 \pm 11$ | $88 \pm 3$ | $29 \pm 2$ | $-76 \pm 5$ | 13 |
| 273.160 | $119004 \pm 15$ | $877 \pm 14$ | $84 \pm 3$ | $22 \pm 3$ | $-59 \pm 7$ | 8 |
| 300.000 | $130715 \pm 10$ | $1191 \pm 3$ | $86.8 \pm 0.2$ | | | 105 |
| 325.000 | $141549 \pm 12$ | $1498 \pm 3$ | $71.6 \pm 0.2$ | | | 107 |
| 350.000 | $152376 \pm 13$ | $1737 \pm 4$ | $60.4 \pm 0.2$ | | | 92 |
| | | | $(0.90 \, N_2 + 0.10 \, H_2)$ | | | |
| 260.000 | $119124 \pm 13$ | $818 \pm 13$ | $92 \pm 3$ | $22 \pm 3$ | $-57 \pm 8$ | 17 |
| 273.160 | $125115 \pm 16$ | $1026 \pm 14$ | $85 \pm 3$ | $17 \pm 3$ | $-45 \pm 6$ | 6 |
| 300.000 | $137386 \pm 11$ | $1349 \pm 3$ | $83.0 \pm 0.2$ | | | 98 |
| 325.000 | $148770 \pm 13$ | $1635 \pm 4$ | $69.0 \pm 0.2$ | | | 76 |

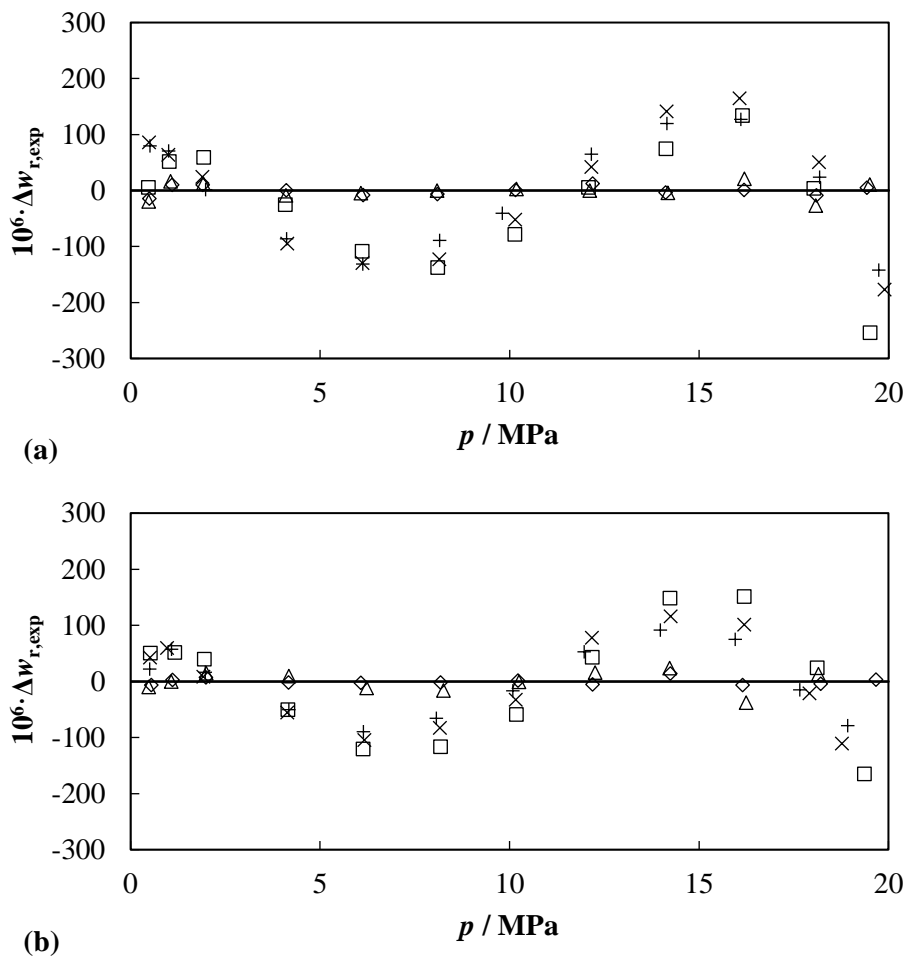

**Figure 5.** Residual plots $\Delta w_{r,exp} = (w_{fitted} - w_{exp})/w_{exp}$ as a function of the pressure for (a) the (0.95 N$_2$ + 0.05 H$_2$) mixture and (b) the (0.90 N$_2$ + 0.10 H$_2$). $w_{fitted}$ are calculated with the equation (2). Temperatures $T = \triangle$ 260 K; $\diamondsuit$ 273.16 K; $\square$ 300 K; $\times$ 325 K; $+$ 350 K.

**Table 6.** Adiabatic coefficient $\gamma^{pg}$, acoustic second virial coefficient $\beta_a$, and acoustic third virial coefficient $\gamma_a$ for the ($N_2 + H_2$) mixtures analysed in this work, with their corresponding relative expanded ($k = 2$) uncertainties, and comparison with AGA8-DC92 EoS [18,19], GERG-2008 EoS [11,12], and GERG-$H2\_improved$ [56]. The superscript $pg$ indicates perfect-gas property.

| $T$ / K | $\gamma^{pg}$ | $10^2 \cdot U_r(\gamma^{pg})$ | $10^2 \cdot \Delta\gamma^{pg}_{\text{AGA8-DC92}}{}^{(*)}$ | $10^2 \cdot \Delta\gamma^{pg}_{\text{GERG-2008}}{}^{(*)}$ | $10^2 \cdot \Delta\gamma^{pg}_{\text{GERG-}H2\_improved}{}^{(*)}$ |
|---|---|---|---|---|---|
| | | | $(0.95\ N_2 + 0.05\ H_2)$ | | |
| 260.000 | 1.3996 | 0.02 | −0.045 | −0.058 | −0.058 |
| 273.160 | 1.3997 | 0.02 | −0.025 | −0.033 | −0.033 |
| 300.000 | 1.3999 | 0.02 | 0.011 | 0.0096 | 0.0096 |
| 325.000 | 1.3993 | 0.02 | −0.0011 | −0.0011 | −0.0011 |
| 350.000 | 1.3988 | 0.02 | −0.0035 | −0.0035 | −0.0035 |
| | | | $(0.90\ N_2 + 0.10\ H_2)$ | | |
| 260.000 | 1.4004 | 0.02 | −0.042 | −0.042 | −0.042 |
| 273.160 | 1.4000 | 0.02 | −0.048 | −0.047 | −0.047 |
| 300.000 | 1.3998 | 0.02 | −0.022 | −0.020 | −0.020 |
| 325.000 | 1.3992 | 0.02 | −0.026 | −0.024 | −0.024 |
| 350.000 | 1.3984 | 0.02 | −0.036 | −0.034 | −0.034 |

| $10^8 \cdot \beta_a$ / $(\text{m}^3 \cdot \text{mol}^{-1})$ | $10^2 \cdot U_r(\beta_a)$ | $10^2 \cdot \Delta\beta_{a,\text{AGA8-DC92}}{}^{(*)}$ | $10^2 \cdot \Delta\beta_{a,\text{GERG-2008}}{}^{(*)}$ | $10^2 \cdot \Delta\beta_{a,\text{GERG-}H2\_improved}{}^{(*)}$ | $10^{12} \cdot \gamma_a$ / $(\text{m}^3 \cdot \text{mol}^{-1})^2$ | $10^2 \cdot U_r(\gamma_a)$ | $10^2 \cdot \Delta\gamma_{a,\text{AGA8-DC92}}{}^{(*)}$ | $10^2 \cdot \Delta\gamma_{a,\text{GERG-2008}}{}^{(*)}$ | $10^2 \cdot \Delta\gamma_{a,\text{GERG-}H2\_improved}{}^{(*)}$ |
|---|---|---|---|---|---|---|---|---|---|
| | | | | $(0.95\ N_2 + 0.05\ H_2)$ | | | | | |

| 260.000 | 1294 | 2 | 13.1 | 9.6 | 9.6 | 3485 | 3 | −18 | −16 | −16 |
|---|---|---|---|---|---|---|---|---|---|---|
| 273.160 | 1674 | 2 | 8.7 | 5.7 | 5.7 | 3506 | 4 | −17 | −14 | −14 |
| 300.000 | 2272 | 0.3 | 2.4 | −0.22 | −0.22 | 4066 | 0.2 | −2.3 | 3.3 | 3.3 |
| 325.000 | 2859 | 0.2 | 4.8 | 2.1 | 2.1 | 3725 | 0.3 | −10 | −3.3 | −3.3 |
| 350.000 | 3317 | 0.2 | 5.3 | 2.6 | 2.6 | 3508 | 0.3 | −15 | −7.2 | −7.2 |
| | | | | $(0.90\ N_2 + 0.10\ H_2)$ | | | | | | |
| 260.000 | 1484 | 2 | 10.1 | 6.5 | 6.5 | 3462 | 4 | −14 | −12 | −12 |
| 273.160 | 1863 | 2 | 8.6 | 5.2 | 5.2 | 3377 | 4 | −16 | −13 | −13 |
| 300.000 | 2449 | 0.2 | 4.5 | 1.2 | 1.2 | 3726 | 0.2 | −5.8 | −0.9 | −0.9 |
| 325.000 | 2970 | 0.2 | 5.5 | 2.1 | 2.1 | 3457 | 0.3 | −12 | −5.8 | −5.8 |
| 350.000 | 3387 | 0.2 | 5.6 | 2.2 | 2.2 | 3276 | 0.4 | −17 | −8.9 | −8.9 |

[*] $\Delta X_{EoS} = (X_{exp} - X_{EoS})/X_{EoS}$ with $X = \gamma^{pg}$, $\beta_a$, $\gamma_a$; and EoS = AGA8-DC92 [18,19], GERG-2008 [11,12], GERG-$H_2\_improved$ [12].

## 4. Discussion.

### 4.1 Speed of sound.

The relative deviations of the experimental speed of sound data $w_{\exp}$ determined in this work for the two mixtures of $(N_2 + H_2)$ are obtained by comparison with the calculated values $w_{EoS}$ from the reference Helmholtz energy mixture models AGA8-DC92 EoS [17,18], GERG-2008 EoS [11,12], and GERG-$H_2$_improved EoS [14]. They are listed in Tables 2 and 3 and depicted in Figures 6 and 7. Table 7 presents the average absolute relative deviation, the root mean square of the relative deviations, the relative bias, and the maximum relative deviation:

$$\Delta_{AAD} = \frac{1}{N} \sum_{i=1} \left| \left( \frac{w_{\exp} - w_{EoS}}{w_{EoS}} \right)_i \right|$$

$$\Delta_{RMS} = \frac{1}{N} \sum_{i=1} \left[ \left( \frac{w_{\exp} - w_{EoS}}{w_{EoS}} \right)_i^2 \right]^{1/2}$$

$$\Delta_{Bias} = \frac{1}{N} \sum_{i=1} \left( \frac{w_{\exp} - w_{EoS}}{w_{EoS}} \right)_i$$

$$\Delta_{Max} = \text{Max} \left[ \left( \frac{w_{\exp} - w_{EoS}}{w_{EoS}} \right) \right]$$

(8)

of the experimental speed of sound data compared with the speed of sound given by the aforementioned three reference models in all the mixtures studied in this work.

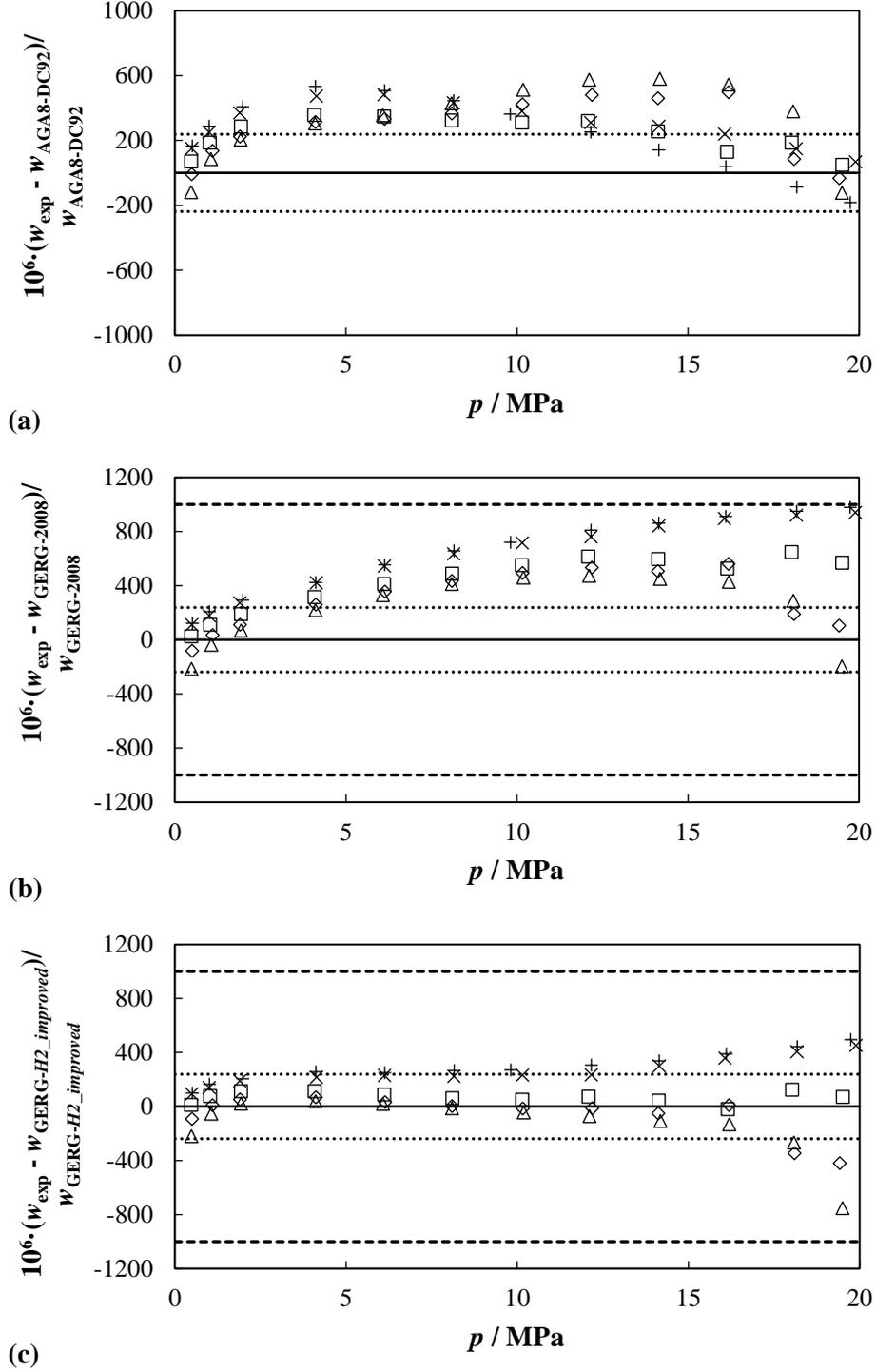

**Figure 6.** (0.95 $N_2$ + 0.05 $H_2$) mixture. Relative deviations upon pressure of experimental speed of sound $\Delta w_{r,EoS} = (w_{exp} - w_{EoS})/w_{EoS}$ with respect to the speed of sound values calculated from (a) AGA8-DC92 EoS [18,19], (b) GERG-2008 EoS [11,12], and (c) GERG-*$H_2$_improved* [12], at temperatures $T = \triangle$ 260 K; $\diamondsuit$ 273.16 K; $\square$ 300 K; $\times$ 325 K; $+$ 350 K. Dotted lines depict the expanded ($k = 2$) uncertainty of the experimental speed of sound; dashed lines depict the expanded ($k = 2$) uncertainty of GERG-*$H_2$_improved* [12].

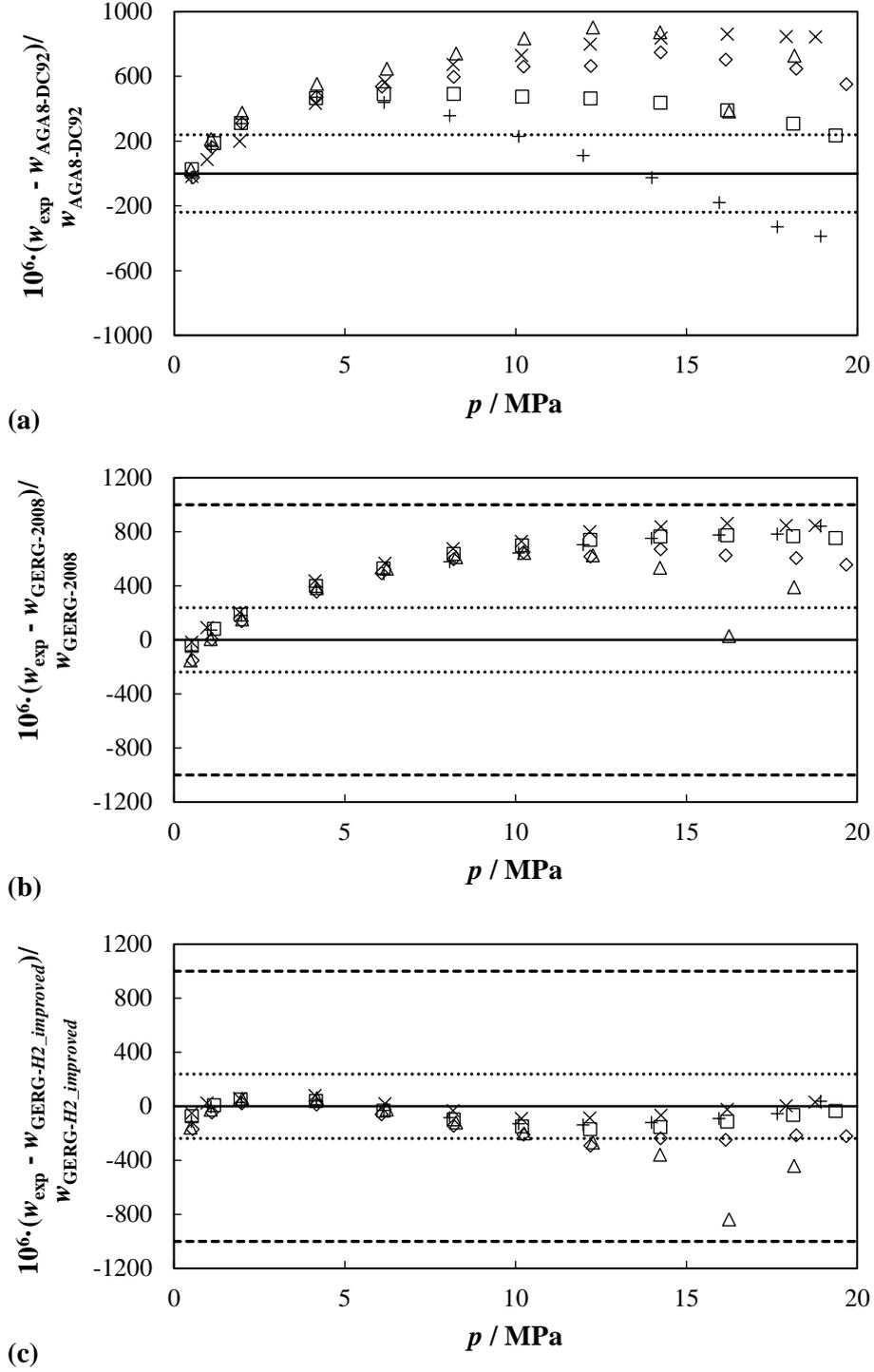

**(a)**

**(b)**

**(c)**

**Figure 7.** $(0.90 \text{ N}_2 + 0.10 \text{ H}_2)$ mixture. Relative deviations upon pressure of experimental speed of sound $\Delta w_{r,\text{EoS}} = (w_{\text{exp}} - w_{\text{EoS}})/w_{\text{EoS}}$ with respect to the speed of sound values calculated from (a) AGA8-DC92 EoS [18,19], (b) GERG-2008 EoS [11,12], and (c) GERG-$H_2$_improved [56], at temperatures $T = \triangle$ 260 K; $\diamond$ 273.16 K; $\square$ 300 K; $\times$ 325 K; $+$ 350 K. Dotted lines depict the expanded ($k = 2$) uncertainty of the experimental speed of sound, dashed lines depict the expanded ($k = 2$) uncertainty of GERG-$H_2$_improved [12].

**Table 7.** Statistical analysis of the speed of sound data with respect to AGA8-DC92 EoS [18,19], GERG-2008 EoS [11,12], and GERG-$H_2\_improved$ [56] for the ($N_2$ + $H_2$) mixtures studied in this work. $\Delta_{AAD}$ = average absolute relative deviations, $\Delta_{Bias}$ = average relative deviations, $\Delta_{RMS}$ = root mean square relative deviations, $\Delta_{Max}$ = maximum relative deviations.

| | $10^2\cdot$(Experimental vs AGA8-DC92) | | | | $10^2\cdot$(Experimental vs GERG-2008) | | | | $10^2\cdot$(Experimental vs GERG-$H_2\_improved$) | | | |
|---|---|---|---|---|---|---|---|---|---|---|---|---|
| | $\Delta_{AAD}$ | $\Delta_{Bias}$ | $\Delta_{RMS}$ | $\Delta_{Max}$ | $\Delta_{AAD}$ | $\Delta_{Bias}$ | $\Delta_{RMS}$ | $\Delta_{Max}$ | $\Delta_{AAD}$ | $\Delta_{Bias}$ | $\Delta_{RMS}$ | $\Delta_{Max}$ |
| (0.95 $N_2$ + 0.05 $H_2$) | 0.029 | 0.027 | 0.033 | 0.058 | 0.045 | 0.043 | 0.052 | 0.098 | 0.017 | 0.008 | 0.023 | 0.075 |
| (0.90 $N_2$ + 0.10 $H_2$) | 0.044 | 0.042 | 0.052 | 0.090 | 0.048 | 0.047 | 0.056 | 0.086 | 0.011 | $-0.010$ | 0.018 | 0.084 |

For the two binary $(0.95\ N_2 + 0.05\ H_2)$ and $(0.90\ N_2 + 0.10\ H_2)$ mixtures, most of the data from an application of the AGA8-DC92 EoS [17,18] and GERG-2008 EoS [11,12] along all the isotherms show positive differences $(w_{exp} - w_{EoS})$ that increase with increasing pressure. The relative discrepancies are within the limit of the expanded $(k = 2)$ experimental uncertainty $U_r(w_{exp}) = 0.022\ \%$ (220 parts in $10^6$) only for pressures above 14 MPa at $T = 350$ K for the AGA8-DC92 EoS [17,18] and at all the isotherms for these two models for pressures below 3 MPa. At equal pressures, the relative deviations with the GERG-2008 EoS [11,12] tend to increase with temperature, but the opposite trend is observed when compared with the AGA8-DC92 EoS [17,18]. The discrepancies are found to be up to +0.090 % and +0.098 % when compared to AGA8-DC92 EoS [17,18] and GERG-2008 EoS [11,12], respectively, but, in any case, they are consistent with these two models, whose stated expanded $(k = 2)$ uncertainties are $U_r(w_{AGA8-DC92})$ = 0.2 % (2000 parts in $10^6$) and $U_r(w_{GERG-2008})$ = 0.5 % (5000 parts in $10^6$). As shown in Table 7, the average of the absolute relative deviations is slightly lower for the AGA8-DC92 EoS [17,18], $\Delta_{AAD} = (0.029\ (x_{H_2} = 0.05)$ and $0.044\ (x_{H_2} = 0.10))$ %, compared with $\Delta_{AAD} = (0.048$ and $0.052)$ % for the GERG-2008 EoS [11,12]. Here, the AGA8-DC92 EoS can reproduce the experimental data for the $(0.95\ N_2 + 0.05\ H_2)$ and $(0.90\ N_2 + 0.10\ H_2)$ mixtures slightly better.

However, the new model GERG-$H_2\_improved$ EoS [14] gives a significantly better result for both mixtures $(x_{H_2} = 0.05$ and $0.10)$ at all conditions investigated. Not only the deviations are within the band of the expanded $(k = 2)$ model uncertainty, $U_r(w_{GERG-H2\_improved}) = 0.1\ \%$ (1000 parts in $10^6$), but rather distributed inside $U_r(w_{exp})$, apart from a few points at pressures above 16 MPa and at $T = (325$ and $350)$ K. $\Delta_{AAD} = (0.017$ and $0.011)$ % are obtained with the model GERG-$H_2\_improved$ EoS [14] which are well below the experimental uncertainty $U_r(w_{exp})$ for the two binary $(0.95\ N_2 + 0.05\ H_2)$ and $(0.90\ N_2 + 0.10\ H_2)$ mixtures. Thus, the development of a specific departure function has really enhanced the performance compared to the original version of the GERG mixing model, allowing to reduce the assigned EoS uncertainty in the speed of sound from 0.5 % to 0.1 %, as demonstrated here.

### 4.2 Second and third acoustic virial coefficients.

Table 6 also lists the relative deviations between the experimental second acoustic virial coefficient $\beta_a$ and third acoustic virial coefficient $\gamma_a$ and the calculated values using AGA8-DC92 EoS [17,18], GERG-2008 EoS [11,12], and GERG-$H_2\_improved$ EoS [14] for the two $(N_2 + H_2)$ mixtures. Experimental $\beta_a$ results show increasing values with increasing both temperature and hydrogen mole fraction that go with a relative expanded $(k = 2)$ uncertainty $U_{r,exp}(\beta_a)$ between (0.2 and 2) %. The differences $(\beta_{a,exp} - \beta_{a,EoS})$ are always positive and one order of magnitude higher than $U_{exp}(\beta_a)$ for all the cases, with the exception of the isotherm at $T = 300$ K for the $(0.95\ N_2 + 0.05\ H_2)$ mixture, where the discrepancies with the GERG-2008 EoS [11,12] and GERG-$H_2\_improved$ EoS [14] are within the uncertainty. The relative deviations with AGA8-DC92 EoS [17,18] are twice those of GERG-2008 EoS [11,12], with the corresponding values of GERG-2008 fairly larger than those of GERG-$H_2\_improved$ EoS [14]. Relative average absolute deviations with the GERG-$H_2\_improved$ EoS [14] are $\Delta_{AAD} = (4.0$ and $3.5)$ % for the $x_{H_2} = (0.05$ and $0.10)$ mixtures respectively.

Data sets of the third acoustic virial coefficient $\gamma_a$ present decreasing values with increasing hydrogen content, for both $(0.95\ N_2 + 0.05\ H_2)$ and $(0.90\ N_2 + 0.10\ H_2)$ mixtures. Within a composition, $\gamma_a$ increases up to $T = 300$ K and slowly decreases for higher temperatures. The relative expanded $(k = 2)$ uncertainties $U_{r,exp}(\gamma_a)$ are between (0.2 and 4) %. Alike the second acoustic coefficient discussed above, the deviations with the three models are at least one order of magnitude higher than the corresponding experimental uncertainty $U_{exp}(\gamma_a)$. But these differences $(\gamma_{a,exp} - \gamma_{a,EoS})$ are now mostly negative, apart from the isotherm at $T = 300$ K for the two mixtures. With this property, AGA8-DC92 EoS [17,18] is still the model rendering poorer predictions of $\gamma_a$ than GERG-2008 EoS [11,12] which in turn performs better than GERG-

*H₂_improved* EoS [14]. Looking at the deviations, that finding is in contrast with the prediction behavior of the models for the second acoustic virial coefficient. Relative average absolute deviations with the GERG-2008 EoS [11,12] are $\Delta_{AAD}$ = (7.0 and 5.0) % for the $x_{H_2}$ = (0.05 and 0.10) mixtures respectively.

## 5. Conclusions.

This research assesses the performance of three established reference thermodynamic models which are based on the Helmholtz energy on binary mixtures of ($N_2 + H_2$) by means of comparing experimental speed of sound data, and acoustic virial coefficients with the results predicted by those reference thermodynamic models. Furthermore, our study provides new and highly accurate experimental data which form the basis for the improvement of these models. A direct benefit of such improvement will be more suitable and less costly designs of the technical components required for the integration of hydrogen into the future energy economy.

Experimental speed of sound data $w(p,T,x)$ for two binary mixtures of nitrogen and hydrogen, with nominal compositions of (0.95 $N_2$ + 0.05 $H_2$) and (0.90 $N_2$ + 0.10 $H_2$), at temperatures between (260 and 350) K and pressures between (0.5 and 20) MPa are reported with an expanded ($k$ = 2) uncertainty of 220 parts in $10^6$ (0.022 %). This leads to a better estimation of the thermodynamic behavior of these mixtures in wide temperature and pressure ranges of interest.

In addition, heat capacity ratios as perfect gas $\gamma^{pg}$, and acoustic virial coefficients $\beta_a$ and $\gamma_a$, have been derived from the speed of sound values. This study also presents a thorough uncertainty analysis of these properties.

The analysis of the measurements reveals: (i) GERG-*H₂_improved* EoS [14] predicts speed of sound data better than both AGA8-DC92 EoS [17,18] and GERG-2008 EoS [11,12] in all cases studied in this work, thus it should be the preferred reference model when dealing with ($N_2 + H_2$) mixtures within the composition, temperature and pressure ranges aforementioned; (ii) GERG-*H₂_improved* EoS [14] is able to reproduce our experimental speed of sound results within the experimental expanded ($k$ = 2) uncertainty for mixtures with hydrogen mole fractions equal to 0.05 and 0.10; and (iii) overall, analogous outcomes applies to the derived properties as perfect-gas $\gamma^{pg}$ and acoustic virial coefficients $\beta_a$ and $\gamma_a$.

## Acknowledgements

The authors want to thank for the support to European Regional Development Fund (ERDF)/Spanish Ministry of Science, Innovation and University (Project ENE2017-88474-R) and ERDF/Regional Government of "Castilla y León" (Project VA280P18). David Vega-Maza is funded by the Spanish Ministry of Science, Innovation and Universities ("Beatriz Galindo Senior" fellowship BEAGAL18/00259).

## References.

[1]    M. De Falco, A. Basile, Enriched Methane, Springer International Publishing, Cham,

       2016. https://doi.org/10.1007/978-3-319-22192-2.

[2]    O. Bičáková, P. Straka, Production of hydrogen from renewable resources and its

       effectiveness, Int. J. Hydrogen Energy. 37 (2012) 11563–11578.

       https://doi.org/10.1016/j.ijhydene.2012.05.047.

[3]    C. Acar, I. Dincer, Comparative assessment of hydrogen production methods from

renewable and non-renewable sources, Int. J. Hydrogen Energy. 39 (2014) 1–12. https://doi.org/10.1016/j.ijhydene.2013.10.060.

[4]     W.J. Nuttall, A.T. Bakenne, Fossil Fuel Hydrogen, Springer International Publishing, Cham, 2020. https://doi.org/10.1007/978-3-030-30908-4.

[5]     K. Altfeld, D. Pinchbeck, Admissible hydrogen concentrations in natural gas systems, Gas Energy. March/2013 (2013) 1–16. https://doi.org/ISSN 2192-158X.

[6]     E.A. Polman, H. de Laat, J. Stappenbelt, P. Peereboom, W. Bouwman, B. de Bruin, C. Pulles, M. Hagen, Reduction of CO2 emissions by adding hydrogen to natural gas. Report no. IE/020726/Pln, Gastec NV, Apeldoorn, The Netherlands, 2003.

[7]     G. Mete, Energy Transitions and the Future of Gas in the EU, Springer International Publishing, Cham, 2020. https://doi.org/10.1007/978-3-030-32614-2.

[8]     J. Gernert, R. Span, EOS-CG: A Helmholtz energy mixture model for humid gases and CCS mixtures, J. Chem. Thermodyn. 93 (2016) 274–293. https://doi.org/10.1016/j.jct.2015.05.015.

[9]     S. Herrig, New Helmholtz-Energy Equations of State for Pure Fluids and CCS-Relevant Mixtures, Ruhr-Universität Bochum, 2018.

[10]    R. Hernández-Gómez, D. Tuma, A. Gómez-Hernández, C.R. Chamorro, Accurate Experimental (p, ρ, T) Data for the Introduction of Hydrogen into the Natural Gas Grid: Thermodynamic Characterization of the Nitrogen-Hydrogen Binary System from 240 to 350 K and Pressures up to 20 MPa, J. Chem. Eng. Data. 62 (2017) 4310–4326. https://doi.org/10.1021/acs.jced.7b00694.

[11]    O. Kunz, R. Klimeck, W. Wagner, M. Jaeschke, GERG Technical Monograph 15 The GERG-2004 wide-range equation of state for natural gases and other mixtures, Düsseldorf, 2007.

[12]    O. Kunz, W. Wagner, The GERG-2008 Wide-Range Equation of State for Natural Gases and Other Mixtures: An Expansion of GERG-2004, J. Chem. Eng. Data. 57 (2012) 3032–3091. https://doi.org/10.1021/je300655b.

[13]    A. Van Itterbeek, W. Van Doninck, Measurements on the velocity of sound in mixtures

of hydrogen, helium, oxygen, nitrogen and carbon monoxide at low temperatures, Proc. Phys. Soc. Sect. B. 62 (1949) 62–69. https://doi.org/10.1088/0370-1301/62/1/308.

[14]    R. Beckmüller, M. Thol, I.H. Bell, E.W. Lemmon, R. Span, New Equations of State for Binary Hydrogen Mixtures Containing Methane , Nitrogen, Carbon Monoxide, and Carbon Dioxide, 013102 (2020) 1–11. https://doi.org/10.1063/5.0040533.

[15]    E.W. Lemmon, I.H. Bell, M.L. Huber, M.O. McLinden, NIST Standard Reference Database 23: Reference Fluid Thermodynamic and Transport Properties-REFPROP, Version 10.0, National Institute of Standards and Technology, (2018) 135. https://doi.org/https://doi.org/10.18434/T4/1502528.

[16]    R.. Span, R.. Beckmüller, S.. Hielscher, A.. Jäger, E.. Mickoleit, T.. Neumann, S.M.. Pohl, B.. Semrau, M. Thol, TREND. Thermodynamic Reference and Engineering Data 5.0, (2020).

[17]    Transmission Measurement Committee, AGA Report No. 8 Part 2 Thermodynamic Properties of Natural Gas and Related Gases GERG–2008 Equation of State, 2017.

[18]    International Organization for Standardization, ISO 20765-1 Natural gas — Calculation of thermodynamic properties — Part 1: Gas phase properties for transmission and distribution applications, Genève, 2005.

[19]    Transmission Measurement Committee, AGA Report No. 8 Part 1 Thermodynamic Properties of Natural Gas and Related Gases DETAIL and GROSS Equations of State, Washington DC, 2017.

[20]    International Organization for Standardization, ISO 6142-1 Gas analysis — Preparation of calibration gas mixtures — Part 1: Gravimetric method for Class I mixtures, Genève, 2014.

[21]    R. Span, E.W. Lemmon, R.T. Jacobsen, W. Wagner, A. Yokozeki, A Reference Equation of State for the Thermodynamic Properties of Nitrogen for Temperatures from 63.151 to 1000 K and Pressures to 2200 MPa, J. Phys. Chem. Ref. Data. 29 (2000) 1361–1433. https://doi.org/10.1063/1.1349047.

[22]    J.W. Leachman, R.T. Jacobsen, S.G. Penoncello, E.W. Lemmon, Fundamental equations

of state for parahydrogen, normal hydrogen, and orthohydrogen, J. Phys. Chem. Ref. Data. 38 (2009) 721–748. https://doi.org/10.1063/1.3160306.

[23]   International Organization for Standardization. ISO 12963, Gas analysis – Comparison methods for the determination of the composition of gas mixtures based on one- and two-point calibration, Genève, 2017.

[24]   D. Lozano-Martín, J.J. Segovia, M.C. Martín, T. Fernández-Vicente, D. del Campo, Speeds of sound for a biogas mixture $CH4 + N2 + CO2 + CO$ from p = (1–12) MPa at T = (273, 300 and 325) K measured with a spherical resonator, J. Chem. Thermodyn. 102 (2016) 348–356. https://doi.org/10.1016/j.jct.2016.07.033.

[25]   J.J. Segovia, D. Lozano-Martín, M.C. Martín, C.R. Chamorro, M.A. Villamañán, E. Pérez, C. García Izquierdo, D. Del Campo, Updated determination of the molar gas constant R by acoustic measurements in argon at UVa-CEM, Metrologia. 54 (2017) 663–673. https://doi.org/10.1088/1681-7575/aa7c47.

[26]   D. Lozano-Martín, A. Rojo, M.C. Martín, D. Vega-Maza, J.J. Segovia, Speeds of sound for $(CH4 + He)$ mixtures from p = (0.5 to 20) MPa at T = (273.16 to 375) K, J. Chem. Thermodyn. 139 (2019). https://doi.org/10.1016/j.jct.2019.07.011.

[27]   D. Lozano-Martín, M.C. Martín, C.R. Chamorro, D. Tuma, J.J. Segovia, Speed of sound for three binary $(CH4 + H2)$ mixtures from p = (0.5 up to 20) MPa at T = (273.16 to 375) K, Int. J. Hydrogen Energy. 45 (2020) 4765–4783. https://doi.org/10.1016/j.ijhydene.2019.12.012.

[28]   M.B. Ewing, J.P.M. Trusler, Speeds of sound in CF 4 between 175 and 300 K measured with a spherical resonator, J. Chem. Phys. 90 (1989) 1106–1115. https://doi.org/10.1063/1.456165.

[29]   J.P.M. Trusler, M. Zarari, The speed of sound and derived thermodynamic properties of methane at temperatures between 275 K and 375 K and pressures up to 10 MPa, J. Chem. Thermodyn. 24 (1992) 973–991. https://doi.org/10.1016/S0021-9614(05)80008-4.

[30]   F.J. Pérez-Sanz, J.J. Segovia, M.C. Martín, D. Del Campo, M.A. Villamañán, Speeds of

sound in (0.95 N2 + 0.05 CO and 0.9 N2 + 0.1 CO) gas mixtures at T = (273 and 325) K and pressure up to 10 MPa, J. Chem. Thermodyn. 79 (2014) 224–229. https://doi.org/10.1016/j.jct.2014.07.022.

[31]     C. Tegeler, R. Span, W. Wagner, A New Equation of State for Argon Covering the Fluid Region for Temperatures From the Melting Line to 700 K at Pressures up to 1000 MPa, J. Phys. Chem. Ref. Data. 28 (1999) 779–850. https://doi.org/10.1063/1.556037.

[32]     J.C. Thermodynamics, D. Lozano-martín, D. Vega-maza, A. Moreau, M.C. Martín, D. Tuma, J.J. Segovia, Speed of sound data , derived perfect-gas heat capacities , and acoustic virial coefficients of a calibration standard natural gas mixture and a low-calorific H 2 -enriched mixture, 158 (2021). https://doi.org/10.1016/j.jct.2021.106434.

[33]     D. Lozano-Martín, R. Susial, P. Hernández, T.E. Fernández-Vicente, M.C. Martín, J.J. Segovia, Speed of sound and phase equilibria for (CO2 + C3H8) mixtures, J. Chem. Thermodyn. 158 (2021). https://doi.org/10.1016/j.jct.2021.106464.

[34]     M.R. Moldover, J.B. Mehl, M. Greenspan, Gas-filled spherical resonators: Theory and experiment, J. Acoust. Soc. Am. 79 (1986) 253–272. https://doi.org/10.1121/1.393566.

[35]     J.B. Mehl, Spherical acoustic resonator: Effects of shell motion, J. Acoust. Soc. Am. 78 (1985) 782–788. https://doi.org/10.1121/1.392448.

[36]     J.P.M. Trusler, Physical Acoustics and Metrology of Fluids, CRC Press, Taylor & Francis Group, 1991.

[37]     K.A. Gillis, H. Lin, M.R. Moldover, Perturbations From Ducts on the Modes of Acoustic Thermometers, J. Res. Natl. Inst. Stand. Technol. 114 (2009) 263. https://doi.org/10.6028/jres.114.019.

[38]     D. Lozano-Martín, M.C. Martín, C.R. Chamorro, D. Tuma, J.J. Segovia, Speed of sound for three binary (CH$_4$ + H$_2$) mixtures from p = (0.5 up to 20) MPa at T = (273.16 to 375) K, Int. J. Hydrogen Energy. 45 (2020). https://doi.org/10.1016/j.ijhydene.2019.12.012.

[39]     D. Lozano-Martín, A. Rojo, M.C. Martín, D. Vega-Maza, J.J. Segovia, Speeds of sound for (CH$_4$ + He) mixtures from p = (0.5 to 20) MPa at T = (273.16 to 375) K,

J. Chem. Thermodyn. 139 (2019). https://doi.org/10.1016/j.jct.2019.07.011.

[40] H.M. Ledbetter, W.F. Weston, E.R. Naimon, Low-temperature elastic properties of four austenitic stainless steels, J. Appl. Phys. 46 (1975) 3855–3860. https://doi.org/10.1063/1.322182.

[41] Cryogenic technologies group, Material properties: 304 stainless (UNS S30400), (2020). https://trc.nist.gov/cryogenics/materials/304Stainless/304Stainless_rev.htm.

[42] CryoSoft, Solid materials database, (2020). https://supermagnet.sourceforge.io/solids/MetallicAlloys/SS304/rep/AISI304.pdf.

[43] M.F. Costa Gomes, J.P.M. Trusler, The speed of sound in nitrogen at temperatures between T = 250K and T = 350K and at pressures up to 30 MPa, J. Chem. Thermodyn. 30 (1998) 527–534. https://doi.org/10.1006/jcht.1997.0331.

[44] K. Meier, S. Kabelac, Speed-of-sound measurements in compressed nitrogen and dry air, J. Chem. Eng. Data. 61 (2016) 3941–3951. https://doi.org/10.1021/acs.jced.6b00720.

[45] D. Güsewell, F. Schmeissner, J. Schmid, Density and sound velocity of saturated liquid neon-hydrogen and neon-deuterium mixtures between 25 and 31 K, Cryogenics (Guildf). 10 (1970) 150–154. https://doi.org/10.1016/0011-2275(70)90087-1.

[46] D.H. Liebenberg, Thermodynamic Properties of Fluid n-H 2 in the, (1977).

[47] A. Van Itterbeek, W. Van Dael, A. Cops, Velocity of ultrasonic waves in liquid normal and para hydrogen, Physica. 27 (1961) 111–116. https://doi.org/10.1016/0031-8914(61)90026-X.

[48] K. Matsuishi, E. Gregoryanz, H. Mao, R.J. Hemley, Equation of state and intermolecular interactions in fluid hydrogen from Brillouin scattering at high pressures and temperatures, J. Chem. Phys. 118 (2003) 10683–10695. https://doi.org/10.1063/1.1575196.

[49] D.H. Liebenberg, R.L. Mills, J.C. Bronson, Measurement of $P$ , $V$ , $T$ , and sound velocity across the melting curve of $n$ - $math display$, Phys. Rev. B. 18 (1978)

4526–4532. https://doi.org/10.1103/PhysRevB.18.4526.

[50]   W. Van Dael, A. Van Itterbeek, A. Cops, J. Thoen, Velocity of sound in liquid hydrogen, Cryogenics (Guildf). 5 (1965) 207–212. https://doi.org/10.1016/0011-2275(65)90059-7.

[51]   K. Matsuishi, E. Gregoryanz, H.K. Mao, R.J. Hemley, Equation of state and intermolecular interactions in fluid hydrogen from Brillouin scattering at high pressures and temperatures, J. Chem. Phys. 118 (2003) 10683–10695. https://doi.org/10.1063/1.1575196.

[52]   T. Shimanouchi, Tables of Molecular Vibrational Frequences Consolidated, vol. 1., Natl. Bur. Stand. (1972) 1–16. https://doi.org/10.6028/NBS.NSRDS.39.

[53]   Joint Committee for Guides in Metrology, Evaluation of measurement data — Guide to the expression of uncertainty in measurement, 2008.

[54]   Joint Committee for Guides in Metrology, Evaluation of measurement data — Supplement 1 to the "Guide to the expression of uncertainty in measurement" — Propagation of distributions using a Monte Carlo method, 2008.

[55]   J. Fischer, B. Fellmuth, C. Gaiser, T. Zandt, L. Pitre, F. Sparasci, M.D. Plimmer, M. De Podesta, R. Underwood, G. Sutton, G. Machin, R.M. Gavioso, D. Madonna Ripa, P.P.M. Steur, J. Qu, X.J. Feng, J. Zhang, M.R. Moldover, S.P. Benz, D.R. White, L. Gianfrani, A. Castrillo, L. Moretti, B. Darquié, E. Moufarej, C. Daussy, S. Briaudeau, O. Kozlova, L. Risegari, J.J. Segovia, M.C. Martin, D. Del Campo, The Boltzmann project, Metrologia. 55 (2018) R1–R20. https://doi.org/10.1088/1681-7575/aaa790.

[56]   R. Beckmüller, M. Thol, I.H. Bell, E.W. Lemmon, R. Span, New Equations of State for Binary Hydrogen Mixtures Containing Methane , Nitrogen, Carbon Monoxide, and Carbon Dioxide, (2020) 1–11.